\RequirePackage{fix-cm}
\documentclass[smallextended]{svjour3}       
\smartqed  
\usepackage{graphicx}
\usepackage{natbib}

\begin{document}

\title{Ancillary science with {\sc Ariel}:
}
\subtitle{Feasibility and scientific potential of young {stellar object} observations}

\titlerunning{Feasibility of young star observations with {\sc Ariel}}       

\author{Gy\H ur\H us, B.$^{1,2}$  \and Kiss, Cs.$^{1,3}$ \and Morales, J.C.$^{4,5}$ 
\and Nakhjiri, N.$^{4,5}$, \and Marton, G.$^{1}$ \and \'Abrah\'am, P.$^{1,3}$
\and K\'osp\'al, \'A.$^{1,3,6}$, \and Mo\'or, A.$^{1}$ \and Szab\'o, Gy.M.$^{7,8}$
\and Szab\'o, R.$^{1,3,9}$}

\authorrunning{Gy\H ur\H us, B., et al.} 

\institute{\and
$^1$Konkoly Observatory, Research Centre for Astronomy and Earth Sciences, Konkoly Thege 15-17, H-1121 Budapest, Hungary\\
\and
$^2$Imperial College London, London, UK\\
\and
$^3$ELTE Eötvös Loránd University, Institute of Physics, P\'azm\'any P\'eter s\'et\'any 1/A, 1117 Budapest, Hungary\\
\and
$^4$Institut de Ci\'encies de l'Espai (ICE, CSIC), Campus UAB, C/ de Can Magrans s/n, E-08193 Ballaterra, Spain\\
\and 
$^5$Institut d'Estudis Espacials de Catalunya (IEEC), C/ Gran Capita 2-4, E-08034 Barcelona, Spain\\
\and
$^6$Max Planck Institute for Astronomy, K\"onigstuhl 17, 69117 Heidelberg, Germany\\
\and
$^7$ELTE E\"otv\"os Lor\'and University, Gothard Astrophysical Observatory, 9700 Szombathely, Szent Imre h. u. 112, Hungary\\
\and
$^8$MTA-ELTE Exoplanet Research Group, 9700 Szombathely, Szent Imre h. u. 112, Hungary\\
\and
$^9$MTA CSFK Lend\"ulet Near-Field Cosmology Research Group}

\date{Received: date / Accepted: date}

\maketitle

\begin{abstract}
To investigate the feasibility of ancillary target observations with ESA's {\sc Ariel} mission, we compiled a list of potentially interesting young stars: FUors, systems harbouring extreme debris discs and a larger sample of young stellar objects showing strong near/mid-infrared excess. These objects can be observed as additional targets in the waiting times between the scheduled exoplanet transit and occultation observations. After analyzing the schedule for {\sc Ariel} an algorithm was constructed to find the optimal target to be observed in each gap. The selection was mainly based on the slew and stabilization time needed to observe the selected YSO, but it also incorporated the scientific importance of the targets and whether they have already been sufficiently measured. After acquiring an adequately large sample of simulation data, it was concluded that approximately 99.2 \% of the available -- at least one hour long -- gaps could be used effectively. With an average slewing and stabilization time of about 16.7 minutes between scheduled exoplanet transits and ancillary targets, this corresponds to an additional $2881 \pm 56$ hours of active data gathering. When this additional time is used to observe our selected 200 ancillary targets, a typical signal-to-noise ratio of $\sim$10$^4$ can be achieved along the whole spectral window covered by {\sc Ariel}.

\end{abstract}


\section{Introduction \label{sect:intro}}

 {\sc Ariel} is ESA’s M4 mission which will carry out a revolutionary infrared spectroscopic survey of transiting exoplanets \citep{Tinetti2018}. The main goal of the mission is to observe exoplanetary systems at the time of planetary transits and eclipses. The fact that these events occur at specific times is the main constraint for {\sc Ariel} mission planning. The long- term mission planning of {\sc Ariel} optimizes the observation schedule considering these transit constraints,
 target completeness and slewing constraints as well \citep{Morales2017,Morales2020}. According to the simulations with this optimization, the efficiency of observations of exoplanets (the primary science targets) reaches 92\%, notably above the requirements. The remaining effective observation time is spent on calibration, slewing, and other housekeeping activities. Due to the fixed times of the observations, however, gaps or waiting periods remain between the actual active periods. Scheduling simulations \citep{Morales2017} estimate ~19\%, 21\%, 23\%, and 27\% of the total time to be waiting time in the 0.5-1.0, 0.5-2.0, 0.5-3.0, and 0.5-4.0-year periods of the mission, respectively, for observations of the mission reference sample (MRS). Allowing extra observations (i.e. observing more transits than the minimum requested for each target to increase signal-to-noise) would decrease the waiting times to $\sim$16\% of the total time for the whole 3.5-year-long mission (0.5-4.0\,yr), which is still a very substantial amount, in total $\sim$4900\,h. 
 %
 %
 According to recent scheduling results \citep{Morales2020}, approximately 1780 of these gaps would be longer than an hour, corresponding to about 3400 hours of waiting time. 
 With efficient scheduling and target selection these waiting times could be filled with valuable observations of ancillary targets and could maximize {\sc Ariel}’s scientific impact.
 
The unique mid-infrared instrumentation of {\sc Ariel} is ideal to study a wide range of stellar phenomena which are difficult to observe from the ground and one has to rely on rarely available space telescope data to characterise these objects through their near and mid-infrared spectra. Many of these targets are directly related to the primary science goals of {\sc Ariel}, taking a snapshot of the evolution of the planetary system at a stage earlier than the scheduled {\sc Ariel} exoplanet transit or eclipse observations.

\section{The significance of observations of the early stages of stellar evolution with {\sc Ariel} \label{sect:scientific_imp}}




Low-mass stars are formed via the gravitational contraction of dense interstellar cores \citep{mckee2007}. The new-born stars are surrounded by a circumstellar disc of gas and dust, which feeds the growing protostar via mass accretion, { and where eventually the planets of the system form \citep{Williams2011}}. The protoplanetary discs usually disperse by the age of 10 million years, and in the debris disc left behind collisions occur between the planetesimals, leading to dust production  \citep{Wyatt2008}. 

The protoplanetary discs around pre-main sequence stars consist largely of gas, with $\sim$1\% dust. This material is significantly processed during the different stages of star formation. The higher density and lower temperature in the disc midplane result in the formation of ice mantles on the surface of the silicate particles. These mantles increase to {the} stickiness of the particles, leading to grain growth. 

The change in density and temperature in the disc midplane can result in the creation of various complex molecules \citep[e.g.][]{Henning2013}. The higher temperature in the vicinity of the protostar will also drive chemical reactions, leading to the formation of organic molecules. Exploring the chemical inventory of the circumstellar environment -- mainly the disc material, but also in the envelope -- is vital to understand the initial conditions for planet formation and predict the composition of new planets. 
The mid-infrared regime, where {\sc Ariel} will work, is ideal to detect the spectroscopic signatures of the molecular content of the circumstellar discs {\citep{Pontoppidan2014,Tinetti2018}, covering the different stages of star formation, from Class 0 to Class II (see also Sect.~4)}.


Space-borne instruments are particularly well suited for such observations as they are not confined to the atmospheric windows but could cover the whole near- and mid-infrared spectral domain. Mid-infrared spectroscopic observations of young stars have been performed by ESA's Infrared Space Observatory, with NASA's Spitzer Space Telescope, and JAXA’s Akari satellite, and also by the PHT-S spectrograph of the Infrared Space Observtory {\citep{Lee2007,Kim2011,Kim2012,Kospal2012}}. 

During their early, pre-main sequence phase, young stellar objects (YSOs) are highly variable, and their variability becomes less and less violent with age. Optical variability has been a long-known defining characteristic of young stars, but the growing amount of multiepoch infrared data shows that these systems also exhibit flux density variations at infrared wavelengths. While the optical variability is caused either by hot or cold stellar spots, or by extinction changes along the line of sight, mid-infrared variability partly reflects changes in the thermal emission of the disc as a response to its varying irradiation by the central star {\citep{Scholz2013}}. Monitoring the variability of YSOs over months or years is therefore an extremely useful diagnostic tool.

A group of pre-main sequence stars is the class of FU Orionis- (FUor) or EX Lupi- (EXor) type young eruptive stars \citep{Audard2014}. Their luminosity bursts have a very strong impact on the circumstellar disc. During the large outburst of EX Lup, changes in the mineralogy of solids and in the molecular abundances were observed \citep{Abraham2009,Banzatti2012}. Theoretical models of the chemical effects of outbursts on the circumstellar environment (the birthplace of the planetary systems) were developed by \citet{Rab2017} and by \citet{Molyarova2018}. {S}tudies predict changes in some molecular abundances, suggesting that FUor and EXor-type outbursts may play an important role in setting the chemical initial conditions for {planet-}formation {\citep{VisserANDBergin2012,Vorobyov2013,Visser2015}}.

Systems harbouring an extreme debris disc (EDDs) represent a special, unusually dust-rich subclass of warm debris discs \citep{Meng2012,Meng2015}. The fractional luminosity (that is the fraction of the stellar luminosity absorbed and re-radiated by the debris dust) of EDDs is higher than 0.01, and their dust temperatures are typically higher than 300\,K. Interestingly, mid-IR photometric monitoring of these objects demonstrated that most of them show significant variability on monthly to yearly timescales. Contrary to typical debris systems, the peculiar dust content of these discs and the observed rapid variations cannot be explained by the steady state collisional evolution of a planetesimal belt (a massive analogue of our asteroid belt). Instead, their observed properties point to a recent, episodic increase in dust production in the inner 1-2\,au region that is thought to be attributed to the final accumulation phase of terrestrial planets \citep{Meng2015}. 

Observations of these subgroups of young stars with near- and mid-infrared spectroscopy, a task that {\sc Ariel} can perform, is a key to understanding the first steps and initial conditions of planet formation. 

\section{Scheduling simulations \label{sect:simulations}}

Planning the observations of transits and eclipses of about 1000 exoplanets is a complicated problem given the large number of possible combinations and the stringent time constraint on such events. {\sc Ariel} will solve this problem using an automatic scheduler based on artificial intelligence algorithms \citep[see][for further details]{Morales2020}, which aim to optimize the mission planning, maximizing both the number of surveyed targets, and the total time used for scientific observations. This scheduling algorithm produces a timeline of tasks by taking into account the list of exoplanets to be observed, the mission constraints and the operations that should be planned. Due to the time constraints of exoplanet transit and occultations, part of the time is lost in gaps of unused time between observations. Typically, these gaps last up to few hours and accumulate to about 4500 hours.

The timeline provided by the simulations \citep{Morales2020} contains the start and end time of the gaps as well as the coordinates of the preceding and subsequent exoplanet transits. This way the slewing and stabilization time required for the ancillary observation can be estimated. Since shorter gaps will mainly be used for calibration purposes and reaction wheel dumping, only the gaps longer than one hour have been considered in our analysis. The total number of these gaps {depends on the actual} launch time, but it averages at about 3400 hours in the 0.5-4.0-year mission duration.

\section{Selection of potential ancillary targets \label{sect:targetselection}}

Provided that the {\sc Ariel} mission will have around 3400 hours worth of useful -- minimum 1 hour long -- gaps over it's 3.5 year long operation, an ancillary target list of 200 targets was constructed
by combining three main sources of stellar information: an available list of known FUors \citep{Connelley2018ANS}, a list of known EDDs
and a pre-compiled list of YSOs. 

The full list of EDDs contains 42 objects, and in all cases the central objects are F-K type stars located within 400\,pc. Of these disc systems 10 were previously known, discovered mainly by the Spitzer Space Telescope. The other 32 objects are new discoveries that we identified using a combined data set based on the AllWISE mid-infrared photometric \citep{cutri2013} and Gaia DR2 astrometric \citep{gaiadr2} catalogs. All EDDs exhibit excess emission already at 4.6\,$\mu$m, most of them even at 3.4\,$\mu$m (Mo\'or et al., 2020, in prep.). 

The available list of FUors and EDDs were used in their entirety due to their short length. To compile the YSO candidate list we used the probabilistic catalog of Gaia+AllWISE YSOs by \citet{Marton2019}. Only those candidates which had at least a 90\% probability of being a YSO were used. {Based on the WISE flux densities we estimate that in our YSO sample $\sim$40\% of the targets are Class 0 or Class I objects, and $\sim$60\% of the targets are Class II objects or transitional discs \citep{Marton2019}, indicating a larger number of objects that could be observed at later evolutionary phases.}

We also considered the distance of the candidates and kept only those with an estimated distance smaller than 1\,kpc, based on the values calculated in \citet{BailerJones2018}. For each object we then computed the healpixel (using nside\,=\,16) in which they are found and selected the brightest candidate (measured in the 2MASS K$_s$ band) in that specific healpixel. 

The final sample had 1552 objects in total. For the present feasibility study this was then reduced by considering their visibility, scientific importance and location, as discussed below. 


In order to determine which targets would be sufficiently bright, a SED comparison with the brightest and faintest stars (HD219134 and GJ1214) that {\sc Ariel} can observe was made \citep{Puig2018}.
Assuming a 5\% uncertainty in the used spectral data, a scientifically more {unique} target {(FUor or EDD)} was selected if it fell between the appropriate intensity interval for at least 40\% of the observable region. A scientifically less unique target was selected if it was sufficiently bright in 50\% of the measurable region. It was found that due to the unusual SED profile of young stellar objects the targets selected were generally better visible on longer wavelengths (mid-infrared region), which is fortunately the more desired region for these objects anyway.


Based on the brightness analysis 18 FUors and 7 EDDs were selected. After the 25 scientifically more relevant targets had been added to the list, the remaining 175 YSOs were selected based on their position in the sky. This way a relatively even distribution of targets was achieved.
In this selection process we assumed that {\sc Ariel} would be targeting every coordinate in the sky with equal probability, which is in fact incorrect, as the orbit of the Earth around the Sun determines which regions of the sky are visible and which are not at a specific date, meaning that targets closer to the ecliptic are only visible for about 40-50\% of the time \citep{Puig2018} and objects closer to the ecliptic poles have an intrinsically higher chance for selection. One may compensate for this effect by giving higher weight for selection of objects closer to the ecliptic, but we did not apply this in our selection process. 


FUors also require a different approach when it comes to observation and scheduling. Due to their unique outbursts, their brightness can change over a few months meaning continuous monitoring would be required to maximize the scientific impact of the ancillary sciences done with {\sc Ariel}. Based on the previous brightness analysis a 1.5 hours long measurement was assigned to each of these targets biannually throughout our simulation.

\section{Simulation \label{sect:simulation}}
\begin{figure}[h!]
\centering
  \includegraphics[width=0.48\textwidth]{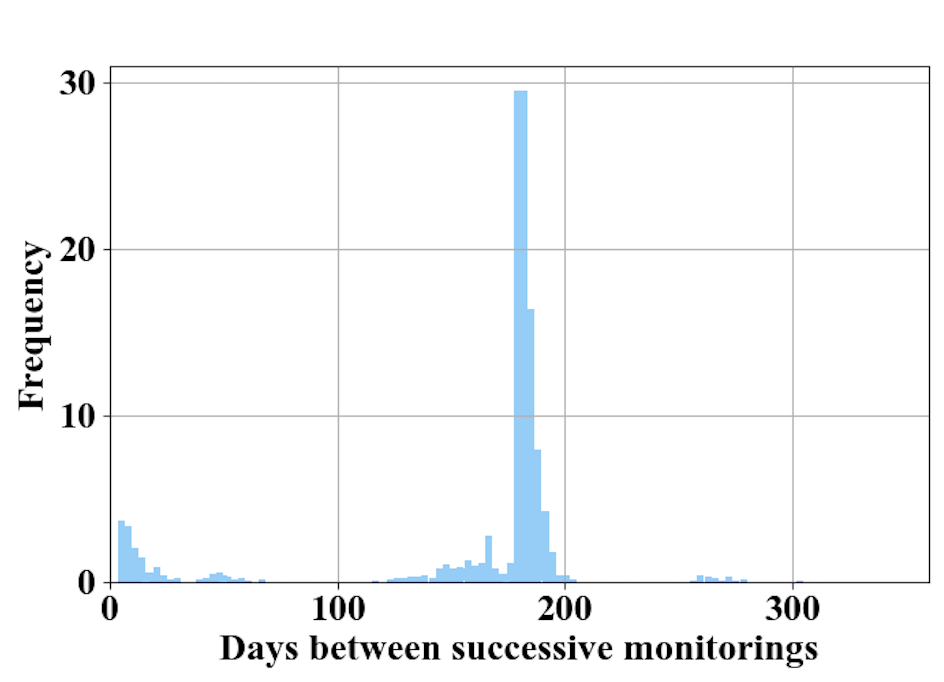}
  \includegraphics[width=0.485\textwidth]{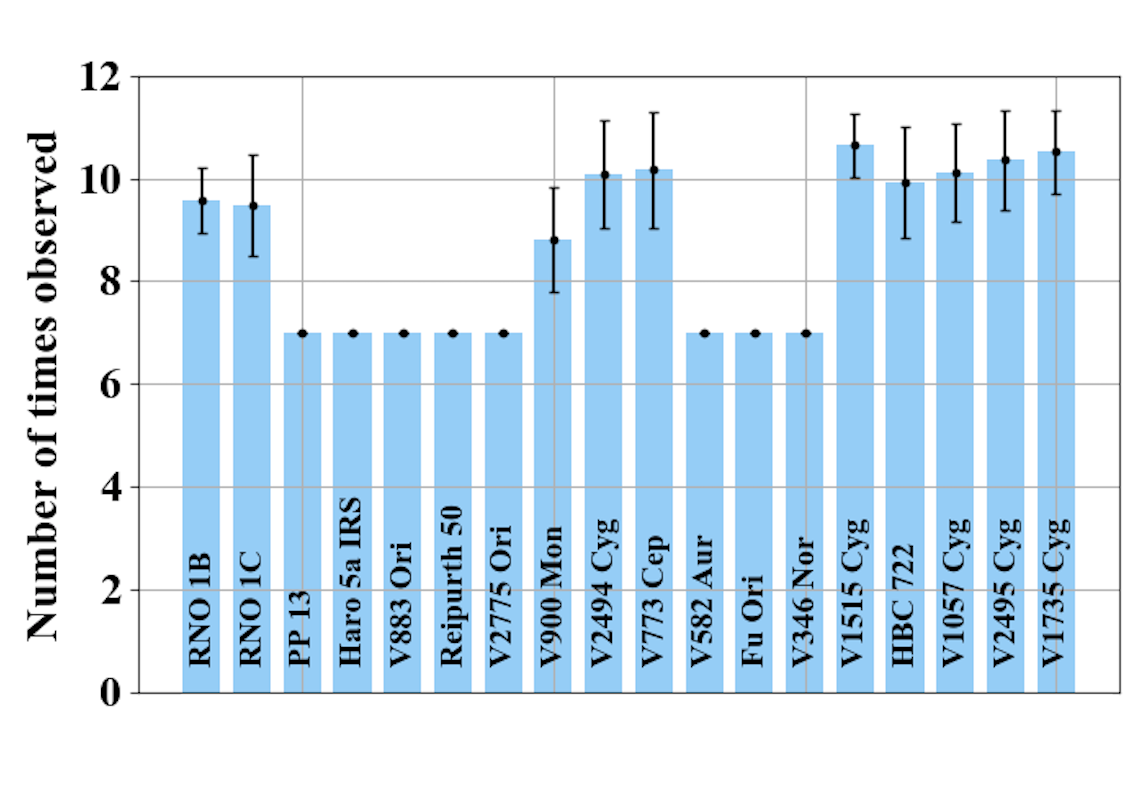}
\caption{The figure shows the distribution of time elapsed between two consecutive measurements of a monitored target (left); and the number of observations of each target selected for monitoring (right).}
\label{fig:monitoring_work_check}
\end{figure}

To investigate the plausibility of observing auxiliary targets using the waiting times, an algorithm was constructed using the received timelines  \citep{Morales2020}. 
At each available gap the algorithm scans through the ancillary targets that are inside the field of view of the device, and assigns a value {($\mu_i$)} to each of them based on the additional slew time and stabilization time required to make the observation, the scientific importance of the target, and the time already spent observing the target. At each gap the algorithm then selects the target that corresponds to the highest value and observes it. As the analysis of targets becomes redundant after a certain number of hours spent measuring them, a limit of 20 hours was set for the total amount of time spent on a YSO. After this the target is marked as if it is no longer visible, even if it is inside the field of view of {\sc Ariel}. These \textit{capped} targets are only selected again, if no other objects are available. This happens if all available targets can only be observed for a maximum of half an hour after the slewing and stabilization. If no visible, capped or uncapped targets are observable for at least 30 minutes the algorithm just disregards the possibility of using the gap effectively. \\
It was found that it is more beneficial to monitor some targets -- e.g. FUors -- throughout the mission time of {\sc Ariel} than to constantly observe them until the limiting 20 hours of observation time is reached. The way it was achieved in the algorithm is that the {aforementioned assigned} value {($\mu_i$)} of these targets was highly boosted every 180 days until they were observed, while it was significantly decreased in between. During the 3.5 years of simulated mission time the 18 targets selected for monitoring were observed approximately $8.66 \pm 1.53$ times, while the average time between two consecutive measurements peaked around $161.5 \pm 7.6$ days (also see Fig.~\ref{fig:monitoring_work_check}). 

\section{Results \label{sect:results}}
\begin{figure}[h!]
\centering
  \includegraphics[width=0.49\textwidth]{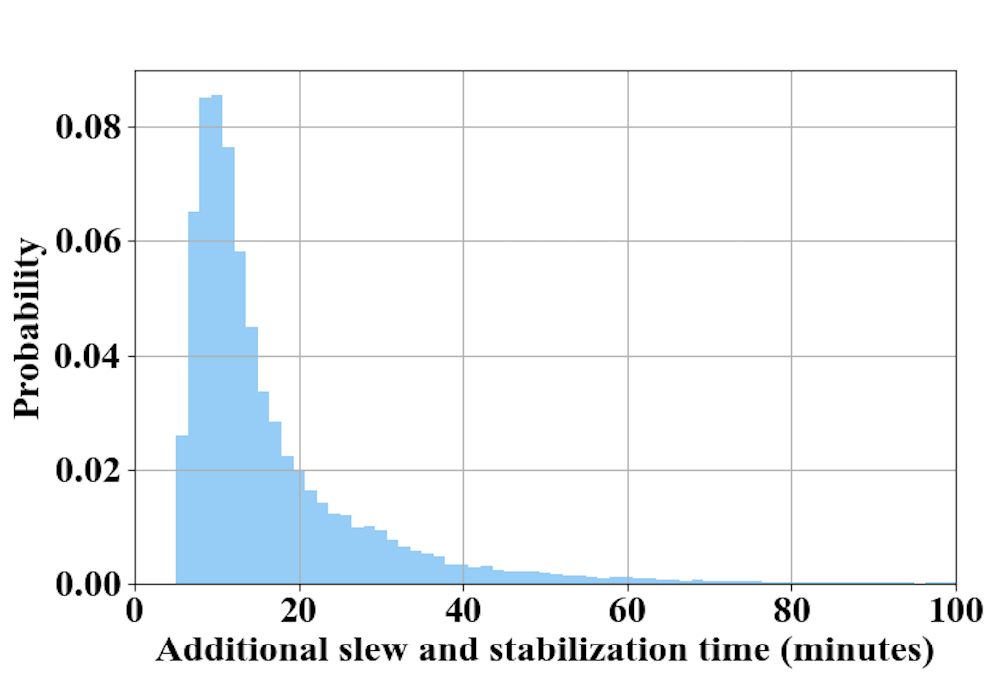}
  \includegraphics[width=0.48\textwidth]{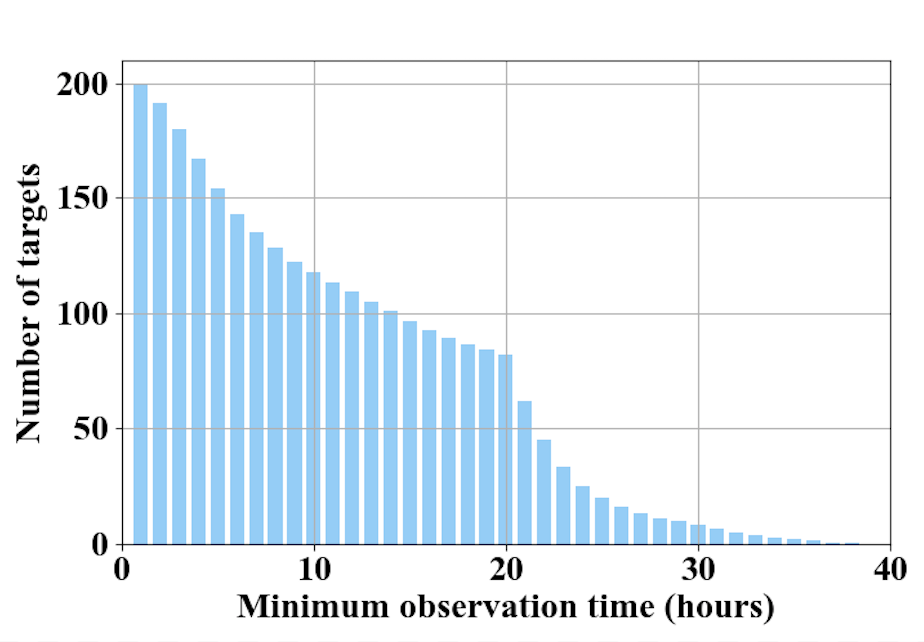}
\caption{These graphs show the general distribution of the slewing and stabilization time additionally required to observe the auxiliary targets (left); and the number of targets as a function of the minimum number of hours they have been observed for (right).}
\label{fig:slew_and_stb}
\end{figure}

To acquire statistically significant results the algorithm was run through 25 different simulations corresponding to 25 potential timelines of {\sc Ariel}. It was found that over its 3.5 years long operation the mission will have $1776.7 \pm 35.9$ gaps that are longer than an hour, corresponding to about $3389.1 \pm 68.8$ hours of free time \citep{Morales2020}. Our results show that 99.16\% of these gaps could be used for auxiliary observations corresponding to an additional $2880.7 \pm 55.6$ hours of active data gathering, without the observational overheads. The average slewing and stabilization time of the auxiliary target observations is 16.7 minutes (see also Fig.~\ref{fig:slew_and_stb} left). Note that the slewing calculation was performed by changing the local longitude and latitude separately in the coordinate system of the spacecraft (with the main axis pointing in the direction of the Sun), which resulted in a small overestimate of the slewing time -- i.e. a small underestimate of the actual observing time spent on the ancillary target. 

\begin{figure}[h!]
\centering
  \includegraphics[width=0.85\textwidth]{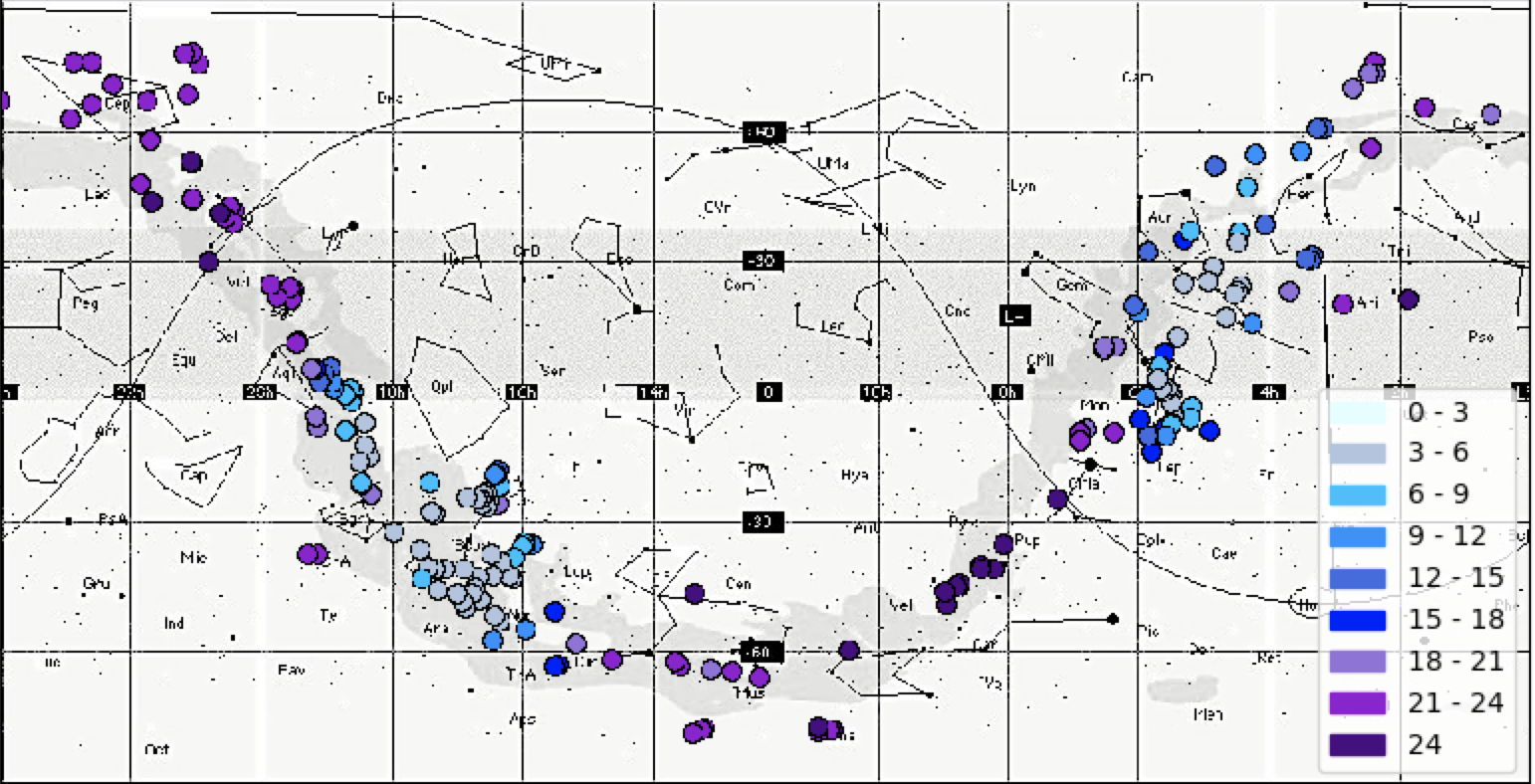}
\caption{The distribution and the visibility of the selected 200 targets after the algorithm has analyzed their potential for ancillary measurements. The colours indicate the hours spent observing the targets, as shown with the colour codes on the right.}
\label{fig:vis_distr}
\end{figure}

Overall it can be deduced that out of the 200 targets selected for analysis, $161 \pm 5$ are observed for at least 5 hours, and $119 \pm 4$ are observed for at least 10 hours (Fig.~\ref{fig:slew_and_stb}, right panel). The visibility and distribution of the targets are shown in Fig.~\ref{fig:vis_distr}.

\section{Signal-to-noise estimates of YSO targets \label{sect:ysosn}}

Another important aspect of this feasibility study is to estimate the signal-to-noise ratio achievable for a specific object during a single waiting time measurement, and with multiple measurements considering the whole mission duration. In this analysis we did not consider the monitoring targets, i.e. FUors and EDDs. 

The spectral energy distribution of our YSO targets were estimated using the Gaia DR2 G \citep{gaiadr2}, 2MASS J, H, K, and WISE W1, W2, W3 magnitudes \citep{cutri2013} by converting them to flux densities (Jy) and interpolating to a fine grid of 102 bins in the 0.5--7.8\,$\mu$m wavelength range. 

\begin{figure}
    \centering
    \includegraphics[width=0.48\textwidth]{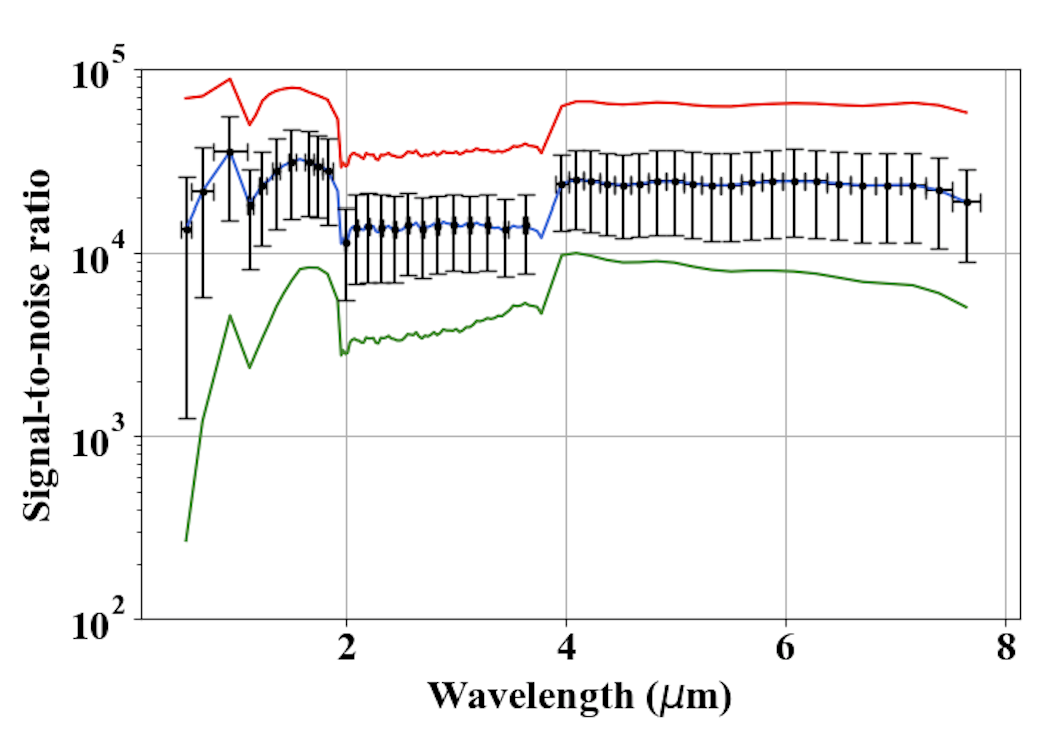}
    \includegraphics[width=0.48\textwidth]{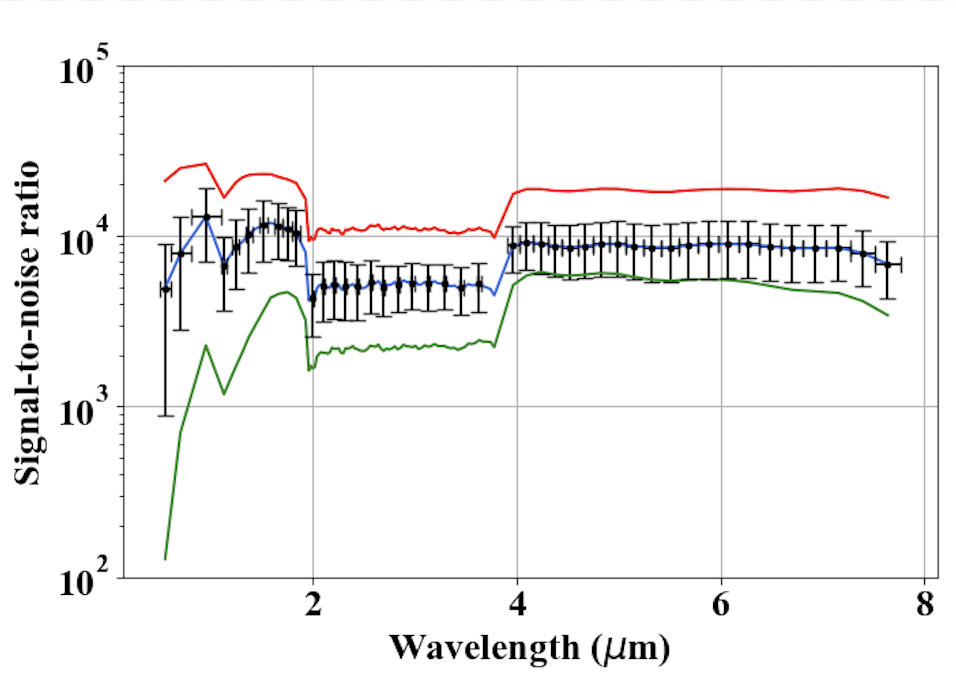}
    \caption{Estimated signal-to-noise ratio of the YSO targets observed in the waiting times, using the sum of all measurements during the mission (left) and in a single measurement with median length (right). The blue curve corresponds to the median signal-to-noise ratio achieved for 175 targets, and the error bars represent the standard deviation within this sample. Red and green curves correspond to the signal-to-noise obtained for the brightest and faintest target, respectively. Note that not all spectral bins are shown.} 
    \label{fig:ysosn}
\end{figure}

To get noise estimates, first we used a reference set of stars of spectral type A-M, placed at distances of 10-300\,pc, and obtained their {\sc Ariel} specific noise estimates using ExoSim \citep{exosim}, for each spectral bin. 
{This provides a reference frame of noise values as a function of brightness in each spectral bin. We used these values to obtain signal-to-noise estimates for our targets, based on the estimated actual brightness. As the range of flux densities provided by the selected A-M-type stars was wide enough it was possible to fully cover the brightness range of our YSOs.}
The results are presented in Fig.~\ref{fig:ysosn}. Note that the signal-to-noise depends on the actual width of the spectral band; in our dataset we used $\Delta\lambda$\,$\approx$\,0.03\,$\mu$m in the 1.95--3.9\,$\mu$m range (AIRS-CH0), and $\Delta\lambda$\,$\approx$\,0.20\,$\mu$m in the 3.9--7.8\,$\mu$m range (AIRS-CH1). Noise estimates are also provided for three photometric bands centered at 0.55, 0.70 and 0.95\,$\mu$m (VISPhot, FGS1, FGS2).  

For our selected YSO targets the typical signal-to-noise in the photometric bands are of the order of 10$^4$ {in the 0.5-2 $\mu$m range}; $\sim$5$\cdot$10$^3$ in the 2--4\,$\mu$m range; and $\sim$10$^4$ in the 4--8\,$\mu$m range in a single measurement. 
{However, these signal-to-noise values are expected to be about three times larger once all the measurements taken over the course of the mission are combined.}

Our results show that a typical YSO selected by our criteria can be observed with a high signal-to-noise ratio even in a single measurement. This also means that multiple monitoring observations {are} useful, and could detect even small-scale (1:1000) flux density variations in the infrared, a unique tool in the characterisation of circumstellar material evolution. 

The high YSO signal-to-noise seen in the near- and mid-infrared is due to the fact that YSOs are much brighter than main sequence stars at these wavelengths as a result of the presence of circumstellar material, if the visible range brightness is the same otherwise. Accordingly, the worst signal-to-noise values are obtained for the visible range photometric bands for YSOs are below or around 10$^3$. 

\section{Conclusion and further work}
To examine the feasibility of ancillary science cases with ESA's {\sc Ariel} mission, we compiled a list of young stars, including  FU Orionis-type variable stars, systems harbouring extreme debris discs and young stellar objects. Using the available exoplanet transit and eclipse scheduling timelines \citep{Morales2020} we constructed an algorithm to find the optimal target to be observed in each -- at least one-hour-long -- gap. The selection was mainly based on the target's celestial distance from the location of the preceding and subsequent scheduled exoplanet transits, but the scientific importance of the target and the overall time spent observing it was also taken into account. Some targets required monitoring biannually rather than continuously; in these cases higher priority was assigned to the targets periodically. We found that 99.2 \% of the gaps can be used effectively, corresponding to an additional $2880.7 \pm 55.6$ hours of active data gathering. \\
This is on its own a very promising result, especially because throughout the simulations some non-negligible underestimates have been made using the non-linear slewing and large overhead. The typical signal-to-noise ratio which could be reached with the spectroscopic instruments (NIRSpec, AIRS-CH0, AIRS-CH1) are in the order of $10^4$. This suggests that a much larger sample of targets (of the order of 500-600) could also be used for ancillary observations, and still keeping the achievable signal-to-noise over 10$^3$ for combined measurements. Also, in the case of the brightest targets, waiting times shorter than one hour could as well be used which were excluded from our present investigation.\\ 
{As the additional active data gathering time was predicted by these simulations using 25 different possible schedules for {\sc Ariel}, it is likely that the algorithm is sufficiently flexible and robust to adapt to the changing target list of the mission while conserving the level of efficiency presented in this paper.}\\
The construction of a larger sample of targets and the correction and fine-tuning of the algorithm will be the next steps in further developing ancillary science cases for ESA's {\sc Ariel} mission.

\begin{acknowledgements}
This research was funded by the K-125015, 2018-2.1.7-UK\_GYAK-2019-00021 and 2020-1.2.1-GYAK-00004 grants of the National Research, Development and Innovation Office (NKFIH), Hungary; the Lend\"ulet Program of the Hungarian Academy of Sciences, project No. LP2018-7/2020 and the MW-Gaia COST Action (CA18104). J.C.M and N.N. acknowledge support from the Spanish Ministry of Economy and Competitiveness (MINECO) and the Fondo Europeo de Desarrollo Regional (FEDER) through grants ESP2016-80435-C2-1-R and and PGC2018-098153-B-C33, as well as the support of the Generalitat de Catalunya/CERCA programme. This project has received funding from the European Research Council (ERC) under the European Union's Horizon 2020 research and innovation programme, under grant agreement No 716155 (SACCRED). We thank our reviewer for the comments and suggestions.

\end{acknowledgements}

%
%

\bibliographystyle{spbasic}      
\bibliography{arielauxsci}   

\begin{thebibliography}{30}
\providecommand{\natexlab}[1]{#1}
\providecommand{\url}[1]{{#1}}
\providecommand{\urlprefix}{URL }
\expandafter\ifx\csname urlstyle\endcsname\relax
  \providecommand{\doi}[1]{DOI~\discretionary{}{}{}#1}\else
  \providecommand{\doi}{DOI~\discretionary{}{}{}\begingroup
  \urlstyle{rm}\Url}\fi
\providecommand{\eprint}[2][]{\url{#2}}

\bibitem[{{{\'A}brah{\'a}m} et~al.(2009){{\'A}brah{\'a}m}, {Juh{\'a}sz},
  {Dullemond}, {K{\'o}sp{\'a}l}, {van Boekel}, {Bouwman}, {Henning},
  {Mo{\'o}r}, {Mosoni}, {Sicilia-Aguilar}, and {Sipos}}]{Abraham2009}
{{\'A}brah{\'a}m} P, {Juh{\'a}sz} A, {Dullemond} CP, {K{\'o}sp{\'a}l} {\'A},
  {van Boekel} R, {Bouwman} J, {Henning} T, {Mo{\'o}r} A, {Mosoni} L,
  {Sicilia-Aguilar} A, {Sipos} N (2009) {Episodic formation of cometary
  material in the outburst of a young Sun-like star}. \nat 459(7244):224--226,
  \doi{10.1038/nature08004}, \eprint{0906.3161}

\bibitem[{{Audard} et~al.(2014){Audard}, {{\'A}brah{\'a}m}, {Dunham}, {Green},
  {Grosso}, {Hamaguchi}, {Kastner}, {K{\'o}sp{\'a}l}, {Lodato}, {Romanova},
  {Skinner}, {Vorobyov}, and {Zhu}}]{Audard2014}
{Audard} M, {{\'A}brah{\'a}m} P, {Dunham} MM, {Green} JD, {Grosso} N,
  {Hamaguchi} K, {Kastner} JH, {K{\'o}sp{\'a}l} {\'A}, {Lodato} G, {Romanova}
  MM, {Skinner} SL, {Vorobyov} EI, {Zhu} Z (2014) {Episodic Accretion in Young
  Stars}. In: {Beuther} H, {Klessen} RS, {Dullemond} CP, {Henning} T (eds)
  Protostars and Planets VI, p 387,
  \doi{10.2458/azu\_uapress\_9780816531240-ch017}, \eprint{1401.3368}

\bibitem[{{Bailer-Jones} et~al.(2018){Bailer-Jones}, {Rybizki}, {Fouesneau},
  {Mantelet}, and {Andrae}}]{BailerJones2018}
{Bailer-Jones} CAL, {Rybizki} J, {Fouesneau} M, {Mantelet} G, {Andrae} R (2018)
  {Estimating Distance from Parallaxes. IV. Distances to 1.33 Billion Stars in
  Gaia Data Release 2}. \aj 156(2):58, \doi{10.3847/1538-3881/aacb21},
  \eprint{1804.10121}

\bibitem[{{Banzatti} et~al.(2012){Banzatti}, {Meyer}, {Bruderer}, {Geers},
  {Pascucci}, {Lahuis}, {Juh{\'a}sz}, {Henning}, and
  {{\'A}brah{\'a}m}}]{Banzatti2012}
{Banzatti} A, {Meyer} MR, {Bruderer} S, {Geers} V, {Pascucci} I, {Lahuis} F,
  {Juh{\'a}sz} A, {Henning} T, {{\'A}brah{\'a}m} P (2012) {EX Lupi from
  Quiescence to Outburst: Exploring the LTE Approach in Modeling Blended
  H$_{2}$O and OH Mid-infrared Emission}. \apj 745(1):90,
  \doi{10.1088/0004-637X/745/1/90}, \eprint{1111.2697}

\bibitem[{Connelley and Reipurth(2018)}]{Connelley2018ANS}
Connelley M, Reipurth B (2018) A near-infrared spectroscopic survey of fu
  orionis objects. The Astrophysical Journal 861:145

\bibitem[{{Cutri} et~al.(2013){Cutri}, {Wright}, {Conrow}, {Fowler},
  {Eisenhardt}, {Grillmair}, {Kirkpatrick}, {Masci}, {McCallon}, {Wheelock},
  {Fajardo-Acosta}, {Yan}, {Benford}, {Harbut}, {Jarrett}, {Lake}, {Leisawitz},
  {Ressler}, {Stanford}, {Tsai}, {Liu}, {Helou}, {Mainzer}, {Gettings},
  {Gonzalez}, {Hoffman}, {Marsh}, {Padgett}, {Skrutskie}, {Beck}, {Papin}, and
  {Wittman}}]{cutri2013}
{Cutri} RM, {Wright} EL, {Conrow} T, {Fowler} JW, {Eisenhardt} PRM, {Grillmair}
  C, {Kirkpatrick} JD, {Masci} F, {McCallon} HL, {Wheelock} SL,
  {Fajardo-Acosta} S, {Yan} L, {Benford} D, {Harbut} M, {Jarrett} T, {Lake} S,
  {Leisawitz} D, {Ressler} ME, {Stanford} SA, {Tsai} CW, {Liu} F, {Helou} G,
  {Mainzer} A, {Gettings} D, {Gonzalez} A, {Hoffman} D, {Marsh} KA, {Padgett}
  D, {Skrutskie} MF, {Beck} RP, {Papin} M, {Wittman} M (2013) {Explanatory
  Supplement to the AllWISE Data Release Products}. Explanatory Supplement to
  the AllWISE Data Release Products

\bibitem[{{Gaia Collaboration} et~al.(2018){Gaia Collaboration}, {Brown},
  {Vallenari}, {Prusti}, {de Bruijne}, {Babusiaux}, {Bailer-Jones}, {Biermann},
  {Evans}, {Eyer}, {Jansen}, {Jordi}, {Klioner}, {Lammers}, {Lindegren},
  {Luri}, {Mignard}, {Panem}, {Pourbaix}, {Randich}, {Sartoretti}, {Siddiqui},
  {Soubiran}, {van Leeuwen}, {Walton}, {Arenou}, {Bastian}, {Cropper},
  {Drimmel}, {Katz}, {Lattanzi}, {Bakker}, {Cacciari}, {Casta{\~n}eda},
  {Chaoul}, {Cheek}, {De Angeli}, {Fabricius}, {Guerra}, {Holl}, {Masana},
  {Messineo}, {Mowlavi}, {Nienartowicz}, {Panuzzo}, {Portell}, {Riello},
  {Seabroke}, {Tanga}, {Th{\'e}venin}, {Gracia-Abril}, {Comoretto},
  {Garcia-Reinaldos}, {Teyssier}, {Altmann}, {Andrae}, {Audard},
  {Bellas-Velidis}, {Benson}, {Berthier}, {Blomme}, {Burgess}, {Busso},
  {Carry}, {Cellino}, {Clementini}, {Clotet}, {Creevey}, {Davidson}, {De
  Ridder}, {Delchambre}, {Dell'Oro}, {Ducourant},
  {Fern{\'a}ndez-Hern{\'a}ndez}, {Fouesneau}, {Fr{\'e}mat}, {Galluccio},
  {Garc{\'\i}a-Torres}, {Gonz{\'a}lez-N{\'u}{\~n}ez}, {Gonz{\'a}lez-Vidal},
  {Gosset}, {Guy}, {Halbwachs}, {Hambly}, {Harrison}, {Hern{\'a}ndez},
  {Hestroffer}, {Hodgkin}, {Hutton}, {Jasniewicz}, {Jean-Antoine-Piccolo},
  {Jordan}, {Korn}, {Krone-Martins}, {Lanzafame}, {Lebzelter}, {L{\"o}ffler},
  {Manteiga}, {Marrese}, {Mart{\'\i}n-Fleitas}, {Moitinho}, {Mora}, {Muinonen},
  {Osinde}, {Pancino}, {Pauwels}, {Petit}, {Recio-Blanco}, {Richards},
  {Rimoldini}, {Robin}, {Sarro}, {Siopis}, {Smith}, {Sozzetti}, {S{\"u}veges},
  {Torra}, {van Reeven}, {Abbas}, {Abreu Aramburu}, {Accart}, {Aerts},
  {Altavilla}, {{\'A}lvarez}, {Alvarez}, {Alves}, {Anderson}, {Andrei},
  {Anglada Varela}, {Antiche}, {Antoja}, {Arcay}, {Astraatmadja}, {Bach},
  {Baker}, {Balaguer-N{\'u}{\~n}ez}, {Balm}, {Barache}, {Barata}, {Barbato},
  {Barblan}, {Barklem}, {Barrado}, {Barros}, {Barstow}, {Bartholom{\'e}
  Mu{\~n}oz}, {Bassilana}, {Becciani}, {Bellazzini}, {Berihuete}, {Bertone},
  {Bianchi}, {Bienaym{\'e}}, {Blanco-Cuaresma}, {Boch}, {Boeche}, {Bombrun},
  {Borrachero}, {Bossini}, {Bouquillon}, {Bourda}, {Bragaglia}, {Bramante},
  {Breddels}, {Bressan}, {Brouillet}, {Br{\"u}semeister}, {Brugaletta},
  {Bucciarelli}, {Burlacu}, {Busonero}, {Butkevich}, {Buzzi}, {Caffau},
  {Cancelliere}, {Cannizzaro}, {Cantat-Gaudin}, {Carballo}, {Carlucci},
  {Carrasco}, {Casamiquela}, {Castellani}, {Castro-Ginard}, {Charlot},
  {Chemin}, {Chiavassa}, {Cocozza}, {Costigan}, {Cowell}, {Crifo}, {Crosta},
  {Crowley}, {Cuypers}, {Dafonte}, {Damerdji}, {Dapergolas}, {David}, {David},
  {de Laverny}, {De Luise}, {De March}, {de Martino}, {de Souza}, {de Torres},
  {Debosscher}, {del Pozo}, {Delbo}, {Delgado}, {Delgado}, {Di Matteo},
  {Diakite}, {Diener}, {Distefano}, {Dolding}, {Drazinos}, {Dur{\'a}n},
  {Edvardsson}, {Enke}, {Eriksson}, {Esquej}, {Eynard Bontemps}, {Fabre},
  {Fabrizio}, {Faigler}, {Falc{\~a}o}, {Farr{\`a}s Casas}, {Federici},
  {Fedorets}, {Fernique}, {Figueras}, {Filippi}, {Findeisen}, {Fonti},
  {Fraile}, {Fraser}, {Fr{\'e}zouls}, {Gai}, {Galleti}, {Garabato},
  {Garc{\'\i}a-Sedano}, {Garofalo}, {Garralda}, {Gavel}, {Gavras}, {Gerssen},
  {Geyer}, {Giacobbe}, {Gilmore}, {Girona}, {Giuffrida}, {Glass}, {Gomes},
  {Granvik}, {Gueguen}, {Guerrier}, {Guiraud}, {Guti{\'e}rrez-S{\'a}nchez},
  {Haigron}, {Hatzidimitriou}, {Hauser}, {Haywood}, {Heiter}, {Helmi}, {Heu},
  {Hilger}, {Hobbs}, {Hofmann}, {Holland}, {Huckle}, {Hypki}, {Icardi},
  {Jan{\ss}en}, {Jevardat de Fombelle}, {Jonker}, {Juh{\'a}sz}, {Julbe},
  {Karampelas}, {Kewley}, {Klar}, {Kochoska}, {Kohley}, {Kolenberg},
  {Kontizas}, {Kontizas}, {Koposov}, {Kordopatis}, {Kostrzewa-Rutkowska},
  {Koubsky}, {Lambert}, {Lanza}, {Lasne}, {Lavigne}, {Le Fustec}, {Le
  Poncin-Lafitte}, {Lebreton}, {Leccia}, {Leclerc}, {Lecoeur-Taibi},
  {Lenhardt}, {Leroux}, {Liao}, {Licata}, {Lindstr{\o}m}, {Lister}, {Livanou},
  {Lobel}, {L{\'o}pez}, {Managau}, {Mann}, {Mantelet}, {Marchal}, {Marchant},
  {Marconi}, {Marinoni}, {Marschalk{\'o}}, {Marshall}, {Martino}, {Marton},
  {Mary}, {Massari}, {Matijevi{\v{c}}}, {Mazeh}, {McMillan}, {Messina},
  {Michalik}, {Millar}, {Molina}, {Molinaro}, {Moln{\'a}r}, {Montegriffo},
  {Mor}, {Morbidelli}, {Morel}, {Morris}, {Mulone}, {Muraveva}, {Musella},
  {Nelemans}, {Nicastro}, {Noval}, {O'Mullane}, {Ord{\'e}novic},
  {Ord{\'o}{\~n}ez-Blanco}, {Osborne}, {Pagani}, {Pagano}, {Pailler},
  {Palacin}, {Palaversa}, {Panahi}, {Pawlak}, {Piersimoni}, {Pineau}, {Plachy},
  {Plum}, {Poggio}, {Poujoulet}, {Pr{\v{s}}a}, {Pulone}, {Racero}, {Ragaini},
  {Rambaux}, {Ramos-Lerate}, {Regibo}, {Reyl{\'e}}, {Riclet}, {Ripepi}, {Riva},
  {Rivard}, {Rixon}, {Roegiers}, {Roelens}, {Romero-G{\'o}mez}, {Rowell},
  {Royer}, {Ruiz-Dern}, {Sadowski}, {Sagrist{\`a} Sell{\'e}s}, {Sahlmann},
  {Salgado}, {Salguero}, {Sanna}, {Santana-Ros}, {Sarasso}, {Savietto},
  {Schultheis}, {Sciacca}, {Segol}, {Segovia}, {S{\'e}gransan}, {Shih},
  {Siltala}, {Silva}, {Smart}, {Smith}, {Solano}, {Solitro}, {Sordo}, {Soria
  Nieto}, {Souchay}, {Spagna}, {Spoto}, {Stampa}, {Steele},
  {Steidelm{\"u}ller}, {Stephenson}, {Stoev}, {Suess}, {Surdej}, {Szabados},
  {Szegedi-Elek}, {Tapiador}, {Taris}, {Tauran}, {Taylor}, {Teixeira},
  {Terrett}, {Teyssand ier}, {Thuillot}, {Titarenko}, {Torra Clotet}, {Turon},
  {Ulla}, {Utrilla}, {Uzzi}, {Vaillant}, {Valentini}, {Valette}, {van Elteren},
  {Van Hemelryck}, {van Leeuwen}, {Vaschetto}, {Vecchiato}, {Veljanoski},
  {Viala}, {Vicente}, {Vogt}, {von Essen}, {Voss}, {Votruba}, {Voutsinas},
  {Walmsley}, {Weiler}, {Wertz}, {Wevers}, {Wyrzykowski}, {Yoldas},
  {{\v{Z}}erjal}, {Ziaeepour}, {Zorec}, {Zschocke}, {Zucker}, {Zurbach}, and
  {Zwitter}}]{gaiadr2}
{Gaia Collaboration}, {Brown} AGA, {Vallenari} A, {Prusti} T, {de Bruijne} JHJ,
  {Babusiaux} C, {Bailer-Jones} CAL, {Biermann} M, {Evans} DW, {Eyer} L,
  {Jansen} F, {Jordi} C, {Klioner} SA, {Lammers} U, {Lindegren} L, {Luri} X,
  {Mignard} F, {Panem} C, {Pourbaix} D, {Randich} S, {Sartoretti} P, {Siddiqui}
  HI, {Soubiran} C, {van Leeuwen} F, {Walton} NA, {Arenou} F, {Bastian} U,
  {Cropper} M, {Drimmel} R, {Katz} D, {Lattanzi} MG, {Bakker} J, {Cacciari} C,
  {Casta{\~n}eda} J, {Chaoul} L, {Cheek} N, {De Angeli} F, {Fabricius} C,
  {Guerra} R, {Holl} B, {Masana} E, {Messineo} R, {Mowlavi} N, {Nienartowicz}
  K, {Panuzzo} P, {Portell} J, {Riello} M, {Seabroke} GM, {Tanga} P,
  {Th{\'e}venin} F, {Gracia-Abril} G, {Comoretto} G, {Garcia-Reinaldos} M,
  {Teyssier} D, {Altmann} M, {Andrae} R, {Audard} M, {Bellas-Velidis} I,
  {Benson} K, {Berthier} J, {Blomme} R, {Burgess} P, {Busso} G, {Carry} B,
  {Cellino} A, {Clementini} G, {Clotet} M, {Creevey} O, {Davidson} M, {De
  Ridder} J, {Delchambre} L, {Dell'Oro} A, {Ducourant} C,
  {Fern{\'a}ndez-Hern{\'a}ndez} J, {Fouesneau} M, {Fr{\'e}mat} Y, {Galluccio}
  L, {Garc{\'\i}a-Torres} M, {Gonz{\'a}lez-N{\'u}{\~n}ez} J,
  {Gonz{\'a}lez-Vidal} JJ, {Gosset} E, {Guy} LP, {Halbwachs} JL, {Hambly} NC,
  {Harrison} DL, {Hern{\'a}ndez} J, {Hestroffer} D, {Hodgkin} ST, {Hutton} A,
  {Jasniewicz} G, {Jean-Antoine-Piccolo} A, {Jordan} S, {Korn} AJ,
  {Krone-Martins} A, {Lanzafame} AC, {Lebzelter} T, {L{\"o}ffler} W, {Manteiga}
  M, {Marrese} PM, {Mart{\'\i}n-Fleitas} JM, {Moitinho} A, {Mora} A, {Muinonen}
  K, {Osinde} J, {Pancino} E, {Pauwels} T, {Petit} JM, {Recio-Blanco} A,
  {Richards} PJ, {Rimoldini} L, {Robin} AC, {Sarro} LM, {Siopis} C, {Smith} M,
  {Sozzetti} A, {S{\"u}veges} M, {Torra} J, {van Reeven} W, {Abbas} U, {Abreu
  Aramburu} A, {Accart} S, {Aerts} C, {Altavilla} G, {{\'A}lvarez} MA,
  {Alvarez} R, {Alves} J, {Anderson} RI, {Andrei} AH, {Anglada Varela} E,
  {Antiche} E, {Antoja} T, {Arcay} B, {Astraatmadja} TL, {Bach} N, {Baker} SG,
  {Balaguer-N{\'u}{\~n}ez} L, {Balm} P, {Barache} C, {Barata} C, {Barbato} D,
  {Barblan} F, {Barklem} PS, {Barrado} D, {Barros} M, {Barstow} MA,
  {Bartholom{\'e} Mu{\~n}oz} S, {Bassilana} JL, {Becciani} U, {Bellazzini} M,
  {Berihuete} A, {Bertone} S, {Bianchi} L, {Bienaym{\'e}} O, {Blanco-Cuaresma}
  S, {Boch} T, {Boeche} C, {Bombrun} A, {Borrachero} R, {Bossini} D,
  {Bouquillon} S, {Bourda} G, {Bragaglia} A, {Bramante} L, {Breddels} MA,
  {Bressan} A, {Brouillet} N, {Br{\"u}semeister} T, {Brugaletta} E,
  {Bucciarelli} B, {Burlacu} A, {Busonero} D, {Butkevich} AG, {Buzzi} R,
  {Caffau} E, {Cancelliere} R, {Cannizzaro} G, {Cantat-Gaudin} T, {Carballo} R,
  {Carlucci} T, {Carrasco} JM, {Casamiquela} L, {Castellani} M, {Castro-Ginard}
  A, {Charlot} P, {Chemin} L, {Chiavassa} A, {Cocozza} G, {Costigan} G,
  {Cowell} S, {Crifo} F, {Crosta} M, {Crowley} C, {Cuypers} J, {Dafonte} C,
  {Damerdji} Y, {Dapergolas} A, {David} P, {David} M, {de Laverny} P, {De
  Luise} F, {De March} R, {de Martino} D, {de Souza} R, {de Torres} A,
  {Debosscher} J, {del Pozo} E, {Delbo} M, {Delgado} A, {Delgado} HE, {Di
  Matteo} P, {Diakite} S, {Diener} C, {Distefano} E, {Dolding} C, {Drazinos} P,
  {Dur{\'a}n} J, {Edvardsson} B, {Enke} H, {Eriksson} K, {Esquej} P, {Eynard
  Bontemps} G, {Fabre} C, {Fabrizio} M, {Faigler} S, {Falc{\~a}o} AJ,
  {Farr{\`a}s Casas} M, {Federici} L, {Fedorets} G, {Fernique} P, {Figueras} F,
  {Filippi} F, {Findeisen} K, {Fonti} A, {Fraile} E, {Fraser} M, {Fr{\'e}zouls}
  B, {Gai} M, {Galleti} S, {Garabato} D, {Garc{\'\i}a-Sedano} F, {Garofalo} A,
  {Garralda} N, {Gavel} A, {Gavras} P, {Gerssen} J, {Geyer} R, {Giacobbe} P,
  {Gilmore} G, {Girona} S, {Giuffrida} G, {Glass} F, {Gomes} M, {Granvik} M,
  {Gueguen} A, {Guerrier} A, {Guiraud} J, {Guti{\'e}rrez-S{\'a}nchez} R,
  {Haigron} R, {Hatzidimitriou} D, {Hauser} M, {Haywood} M, {Heiter} U, {Helmi}
  A, {Heu} J, {Hilger} T, {Hobbs} D, {Hofmann} W, {Holland} G, {Huckle} HE,
  {Hypki} A, {Icardi} V, {Jan{\ss}en} K, {Jevardat de Fombelle} G, {Jonker} PG,
  {Juh{\'a}sz} {\'A}L, {Julbe} F, {Karampelas} A, {Kewley} A, {Klar} J,
  {Kochoska} A, {Kohley} R, {Kolenberg} K, {Kontizas} M, {Kontizas} E,
  {Koposov} SE, {Kordopatis} G, {Kostrzewa-Rutkowska} Z, {Koubsky} P, {Lambert}
  S, {Lanza} AF, {Lasne} Y, {Lavigne} JB, {Le Fustec} Y, {Le Poncin-Lafitte} C,
  {Lebreton} Y, {Leccia} S, {Leclerc} N, {Lecoeur-Taibi} I, {Lenhardt} H,
  {Leroux} F, {Liao} S, {Licata} E, {Lindstr{\o}m} HEP, {Lister} TA, {Livanou}
  E, {Lobel} A, {L{\'o}pez} M, {Managau} S, {Mann} RG, {Mantelet} G, {Marchal}
  O, {Marchant} JM, {Marconi} M, {Marinoni} S, {Marschalk{\'o}} G, {Marshall}
  DJ, {Martino} M, {Marton} G, {Mary} N, {Massari} D, {Matijevi{\v{c}}} G,
  {Mazeh} T, {McMillan} PJ, {Messina} S, {Michalik} D, {Millar} NR, {Molina} D,
  {Molinaro} R, {Moln{\'a}r} L, {Montegriffo} P, {Mor} R, {Morbidelli} R,
  {Morel} T, {Morris} D, {Mulone} AF, {Muraveva} T, {Musella} I, {Nelemans} G,
  {Nicastro} L, {Noval} L, {O'Mullane} W, {Ord{\'e}novic} C,
  {Ord{\'o}{\~n}ez-Blanco} D, {Osborne} P, {Pagani} C, {Pagano} I, {Pailler} F,
  {Palacin} H, {Palaversa} L, {Panahi} A, {Pawlak} M, {Piersimoni} AM, {Pineau}
  FX, {Plachy} E, {Plum} G, {Poggio} E, {Poujoulet} E, {Pr{\v{s}}a} A, {Pulone}
  L, {Racero} E, {Ragaini} S, {Rambaux} N, {Ramos-Lerate} M, {Regibo} S,
  {Reyl{\'e}} C, {Riclet} F, {Ripepi} V, {Riva} A, {Rivard} A, {Rixon} G,
  {Roegiers} T, {Roelens} M, {Romero-G{\'o}mez} M, {Rowell} N, {Royer} F,
  {Ruiz-Dern} L, {Sadowski} G, {Sagrist{\`a} Sell{\'e}s} T, {Sahlmann} J,
  {Salgado} J, {Salguero} E, {Sanna} N, {Santana-Ros} T, {Sarasso} M,
  {Savietto} H, {Schultheis} M, {Sciacca} E, {Segol} M, {Segovia} JC,
  {S{\'e}gransan} D, {Shih} IC, {Siltala} L, {Silva} AF, {Smart} RL, {Smith}
  KW, {Solano} E, {Solitro} F, {Sordo} R, {Soria Nieto} S, {Souchay} J,
  {Spagna} A, {Spoto} F, {Stampa} U, {Steele} IA, {Steidelm{\"u}ller} H,
  {Stephenson} CA, {Stoev} H, {Suess} FF, {Surdej} J, {Szabados} L,
  {Szegedi-Elek} E, {Tapiador} D, {Taris} F, {Tauran} G, {Taylor} MB,
  {Teixeira} R, {Terrett} D, {Teyssand ier} P, {Thuillot} W, {Titarenko} A,
  {Torra Clotet} F, {Turon} C, {Ulla} A, {Utrilla} E, {Uzzi} S, {Vaillant} M,
  {Valentini} G, {Valette} V, {van Elteren} A, {Van Hemelryck} E, {van Leeuwen}
  M, {Vaschetto} M, {Vecchiato} A, {Veljanoski} J, {Viala} Y, {Vicente} D,
  {Vogt} S, {von Essen} C, {Voss} H, {Votruba} V, {Voutsinas} S, {Walmsley} G,
  {Weiler} M, {Wertz} O, {Wevers} T, {Wyrzykowski} {\L}, {Yoldas} A,
  {{\v{Z}}erjal} M, {Ziaeepour} H, {Zorec} J, {Zschocke} S, {Zucker} S,
  {Zurbach} C, {Zwitter} T (2018) {Gaia Data Release 2. Summary of the contents
  and survey properties}. \aap 616:A1, \doi{10.1051/0004-6361/201833051},
  \eprint{1804.09365}

\bibitem[{{Henning} and {Semenov}(2013)}]{Henning2013}
{Henning} T, {Semenov} D (2013) {Chemistry in Protoplanetary Disks}. Chemical
  Reviews 113(12):9016--9042, \doi{10.1021/cr400128p}, \eprint{1310.3151}

\bibitem[{Kim et~al.(2011)Kim, Evans, Dunham, Chen, Lee, Bourke, Huard,
  Shirley, and Vries}]{Kim2011}
Kim HJ, Evans NJ, Dunham MM, Chen JH, Lee JE, Bourke TL, Huard TL, Shirley YL,
  Vries CD (2011) The spitzer c2d survey of nearby dense cores xi : Infrared
  and submillimeter observations of cb130. The Astrophysical Journal 729(2):84,
  \doi{10.1088/0004-637X/729/2/84}

\bibitem[{Kim et~al.(2012)Kim, Evans, Dunham, Chen, and Pontoppidan}]{Kim2012}
Kim HJ, Evans NJ, Dunham MM, Chen JH, Pontoppidan KM (2012) Co2 ice toward
  low-luminosity embedded protostars: Evidence for episodic mass accretion via
  chemical history. The Astrophysical Journal 758(1):38,
  \doi{10.1088/0004-637X/758/1/38}

\bibitem[{{K{\'o}sp{\'a}l} et~al.(2012){K{\'o}sp{\'a}l}, {{\'A}brah{\'a}m},
  {Acosta-Pulido}, {Dullemond}, {Henning}, {Kun}, {Leinert}, {Mo{\'o}r}, and
  {Turner}}]{Kospal2012}
{K{\'o}sp{\'a}l} {\'A}, {{\'A}brah{\'a}m} P, {Acosta-Pulido} JA, {Dullemond}
  CP, {Henning} T, {Kun} M, {Leinert} C, {Mo{\'o}r} A, {Turner} NJ (2012)
  {Mid-infrared Spectral Variability Atlas of Young Stellar Objects}. \apjs
  201(2):11, \doi{10.1088/0067-0049/201/2/11}, \eprint{1204.3473}

\bibitem[{Lee(2007)}]{Lee2007}
Lee EJ (2007) Chemical evolution in vellos. Journal of the Korean Astronomical
  Society 40:83, \doi{10.5303/JKAS.2007.40.4.083}

\bibitem[{{Marton} et~al.(2019){Marton}, {{\'A}brah{\'a}m}, {Szegedi-Elek},
  {Varga}, {Kun}, {K{\'o}sp{\'a}l}, {Varga-Vereb{\'e}lyi}, {Hodgkin},
  {Szabados}, {Beck}, and {Kiss}}]{Marton2019}
{Marton} G, {{\'A}brah{\'a}m} P, {Szegedi-Elek} E, {Varga} J, {Kun} M,
  {K{\'o}sp{\'a}l} {\'A}, {Varga-Vereb{\'e}lyi} E, {Hodgkin} S, {Szabados} L,
  {Beck} R, {Kiss} C (2019) {Identification of Young Stellar Object candidates
  in the Gaia DR2 x AllWISE catalogue with machine learning methods}. \mnras
  487(2):2522--2537, \doi{10.1093/mnras/stz1301}, \eprint{1905.03063}

\bibitem[{{McKee} and {Ostriker}(2007)}]{mckee2007}
{McKee} CF, {Ostriker} EC (2007) {Theory of Star Formation}. \araa
  45(1):565--687, \doi{10.1146/annurev.astro.45.051806.110602},
  \eprint{0707.3514}

\bibitem[{{Meng} et~al.(2012){Meng}, {Rieke}, {Su}, {Ivanov}, {Vanzi}, and
  {Rujopakarn}}]{Meng2012}
{Meng} HYA, {Rieke} GH, {Su} KYL, {Ivanov} VD, {Vanzi} L, {Rujopakarn} W (2012)
  {Variability of the Infrared Excess of Extreme Debris Disks}. \apjl
  751(1):L17, \doi{10.1088/2041-8205/751/1/L17}, \eprint{1205.1040}

\bibitem[{{Meng} et~al.(2015){Meng}, {Su}, {Rieke}, {Rujopakarn}, {Myers},
  {Cook}, {Erdelyi}, {Maloney}, {McMath}, {Persha}, {Poshyachinda}, and
  {Reichart}}]{Meng2015}
{Meng} HYA, {Su} KYL, {Rieke} GH, {Rujopakarn} W, {Myers} G, {Cook} M,
  {Erdelyi} E, {Maloney} C, {McMath} J, {Persha} G, {Poshyachinda} S,
  {Reichart} DE (2015) {Planetary Collisions Outside the Solar System: Time
  Domain Characterization of Extreme Debris Disks}. \apj 805(1):77,
  \doi{10.1088/0004-637X/805/1/77}, \eprint{1503.05610}

\bibitem[{{Molyarova} et~al.(2018){Molyarova}, {Akimkin}, {Semenov},
  {{\'A}brah{\'a}m}, {Henning}, {K{\'o}sp{\'a}l}, {Vorobyov}, and
  {Wiebe}}]{Molyarova2018}
{Molyarova} T, {Akimkin} V, {Semenov} D, {{\'A}brah{\'a}m} P, {Henning} T,
  {K{\'o}sp{\'a}l} {\'A}, {Vorobyov} E, {Wiebe} D (2018) {Chemical Signatures
  of the FU Ori Outbursts}. \apj 866(1):46, \doi{10.3847/1538-4357/aadfd9},
  \eprint{1809.01925}

\bibitem[{{Morales} et~al.(2017){Morales}, {Garcia-Piquer}, I.~{Ribas},
  {Colom\'e}, {Beaulieu}, {Moneti}, {Coudé du Foresto}, {Duong}, {Queyrel},
  {Jaubert}, {Clédassou}, and {Van-Troostenbergh}}]{Morales2017}
{Morales} JC, {Garcia-Piquer} A, I~{Ribas} I, {Colom\'e} J, {Beaulieu} JP,
  {Moneti} A, {Coudé du Foresto} V, {Duong} B, {Queyrel} J, {Jaubert} J,
  {Clédassou} R, {Van-Troostenbergh} P (2017) {ARIEL Long Term Planning}.
  ARIEL-ICE-GS-TN-001, Issue 1.0

\bibitem[{{Morales} et~al.(2020){Morales}, {Nakhjiri}, {Colom\'e}, {Ribas},
  {Garcia}, {Moreno}, and Vilardell}]{Morales2020}
{Morales} JC, {Nakhjiri} N, {Colom\'e} J, {Ribas} I, {Garcia} E, {Moreno} D,
  Vilardell F (2020) {Ariel scheduling using Artificial Intelligence}.
  Experimental Astronomy, submitted

\bibitem[{Pontoppidan and Blevins(2014)}]{Pontoppidan2014}
Pontoppidan KM, Blevins SM (2014) The chemistry of planet-forming regions is
  not interstellar. Faraday Discussion 168:49--60, \doi{10.1039/C3FD00141E}

\bibitem[{{Puig} et~al.(2018){Puig}, {Pilbratt}, {Heske}, {Escudero},
  {Crouzet}, {de Vogeleer}, {Symonds}, {Kohley}, {Drossart}, {Eccleston},
  {Hartogh}, {Leconte}, {Micela}, {Ollivier}, {Tinetti}, {Turrini},
  {Vandenbussche}, and {Wolkenberg}}]{Puig2018}
{Puig} L, {Pilbratt} G, {Heske} A, {Escudero} I, {Crouzet} PE, {de Vogeleer} B,
  {Symonds} K, {Kohley} R, {Drossart} P, {Eccleston} P, {Hartogh} P, {Leconte}
  J, {Micela} G, {Ollivier} M, {Tinetti} G, {Turrini} D, {Vandenbussche} B,
  {Wolkenberg} P (2018) {The Phase A study of the ESA M4 mission candidate
  ARIEL}. Experimental Astronomy 46(1):211--239,
  \doi{10.1007/s10686-018-9604-3}

\bibitem[{{Rab} et~al.(2017){Rab}, {Elbakyan}, {Vorobyov}, {G{\"u}del},
  {Dionatos}, {Audard}, {Kamp}, {Thi}, {Woitke}, and {Postel}}]{Rab2017}
{Rab} C, {Elbakyan} V, {Vorobyov} E, {G{\"u}del} M, {Dionatos} O, {Audard} M,
  {Kamp} I, {Thi} WF, {Woitke} P, {Postel} A (2017) {The chemistry of episodic
  accretion in embedded objects. 2D radiation thermo-chemical models of the
  post-burst phase}. \aap 604:A15, \doi{10.1051/0004-6361/201730812},
  \eprint{1705.03946}

\bibitem[{{Sarkar} et~al.(2020){Sarkar}, {Pascale}, {Papageorgiou}, {Johnson},
  and {Waldmann}}]{exosim}
{Sarkar} S, {Pascale} E, {Papageorgiou} A, {Johnson} LJ, {Waldmann} I (2020)
  {ExoSim: the Exoplanet Observation Simulator}. arXiv e-prints
  arXiv:2002.03739, \eprint{2002.03739}

\bibitem[{Scholz et~al.(2013)Scholz, Froebrich, and Wood}]{Scholz2013}
Scholz A, Froebrich D, Wood K (2013) A systematic survey for eruptive young
  stellar objects using mid-infrared photometry. Monthly Notices of the Royal
  Astronomical Society 430(4):2910--2922, \doi{10.1093/mnras/stt091}

\bibitem[{{Tinetti} et~al.(2018){Tinetti}, {Drossart}, {Eccleston}, {Hartogh},
  {Heske}, {Leconte}, {Micela}, {Ollivier}, {Pilbratt}, {Puig}, {Turrini},
  {Vandenbussche}, {Wolkenberg}, {Beaulieu}, {Buchave}, {Ferus}, {Griffin},
  {Guedel}, {Justtanont}, {Lagage}, {Machado}, {Malaguti}, {Min},
  {N{\o}rgaard-Nielsen}, {Rataj}, {Ray}, {Ribas}, {Swain}, {Szabo}, {Werner},
  {Barstow}, {Burleigh}, {Cho}, {du Foresto}, {Coustenis}, {Decin}, {Encrenaz},
  {Galand }, {Gillon}, {Helled}, {Morales}, {Mu{\~n}oz}, {Moneti}, {Pagano},
  {Pascale}, {Piccioni}, {Pinfield}, {Sarkar}, {Selsis}, {Tennyson}, {Triaud},
  {Venot}, {Waldmann}, {Waltham}, {Wright}, {Amiaux}, {Augu{\`e}res},
  {Berth{\'e}}, {Bezawada}, {Bishop}, {Bowles}, {Coffey}, {Colom{\'e}},
  {Crook}, {Crouzet}, {Da Peppo}, {Sanz}, {Focardi}, {Frericks}, {Hunt},
  {Kohley}, {Middleton}, {Morgante}, {Ottensamer}, {Pace}, {Pearson},
  {Stamper}, {Symonds}, {Rengel}, {Renotte}, {Ade}, {Affer}, {Alard}, {Allard},
  {Altieri}, {Andr{\'e}}, {Arena}, {Argyriou}, {Aylward}, {Baccani}, {Bakos},
  {Banaszkiewicz}, {Barlow}, {Batista}, {Bellucci}, {Benatti}, {Bernardi},
  {B{\'e}zard}, {Blecka}, {Bolmont}, {Bonfond}, {Bonito}, {Bonomo}, {Brucato},
  {Brun}, {Bryson}, {Bujwan}, {Casewell}, {Charnay}, {Pestellini}, {Chen},
  {Ciaravella}, {Claudi}, {Cl{\'e}dassou}, {Damasso}, {Damiano}, {Danielski},
  {Deroo}, {Di Giorgio}, {Dominik}, {Doublier}, {Doyle}, {Doyon}, {Drummond},
  {Duong}, {Eales}, {Edwards}, {Farina}, {Flaccomio}, {Fletcher}, {Forget},
  {Fossey}, {Fr{\"a}nz}, {Fujii}, {Garc{\'\i}a-Piquer}, {Gear}, {Geoffray},
  {G{\'e}rard}, {Gesa}, {Gomez}, {Graczyk}, {Griffith}, {Grodent}, {Guarcello},
  {Gustin}, {Hamano}, {Hargrave}, {Hello}, {Heng}, {Herrero}, {Hornstrup},
  {Hubert}, {Ida}, {Ikoma}, {Iro}, {Irwin}, {Jarchow}, {Jaubert}, {Jones},
  {Julien}, {Kameda}, {Kerschbaum}, {Kervella}, {Koskinen}, {Krijger}, {Krupp},
  {Lafarga}, {Landini}, {Lellouch}, {Leto}, {Luntzer}, {Rank-L{\"u}ftinger},
  {Maggio}, {Maldonado}, {Maillard}, {Mall}, {Marquette}, {Mathis}, {Maxted},
  {Matsuo}, {Medvedev}, {Miguel}, {Minier}, {Morello}, {Mura}, {Narita},
  {Nascimbeni}, {Nguyen Tong}, {Noce}, {Oliva}, {Palle}, {Palmer}, {Pancrazzi},
  {Papageorgiou}, {Parmentier}, {Perger}, {Petralia}, {Pezzuto},
  {Pierrehumbert}, {Pillitteri}, {Piotto}, {Pisano}, {Prisinzano}, {Radioti},
  {R{\'e}ess}, {Rezac}, {Rocchetto}, {Rosich}, {Sanna}, {Santerne}, {Savini},
  {Scandariato}, {Sicardy}, {Sierra}, {Sindoni}, {Skup}, {Snellen}, {Sobiecki},
  {Soret}, {Sozzetti}, {Stiepen}, {Strugarek}, {Taylor}, {Taylor}, {Terenzi},
  {Tessenyi}, {Tsiaras}, {Tucker}, {Valencia}, {Vasisht}, {Vazan}, {Vilardell},
  {Vinatier}, {Viti}, {Waters}, {Wawer}, {Wawrzaszek}, {Whitworth}, {Yung},
  {Yurchenko}, {Osorio}, {Zellem}, {Zingales}, and {Zwart}}]{Tinetti2018}
{Tinetti} G, {Drossart} P, {Eccleston} P, {Hartogh} P, {Heske} A, {Leconte} J,
  {Micela} G, {Ollivier} M, {Pilbratt} G, {Puig} L, {Turrini} D,
  {Vandenbussche} B, {Wolkenberg} P, {Beaulieu} JP, {Buchave} LA, {Ferus} M,
  {Griffin} M, {Guedel} M, {Justtanont} K, {Lagage} PO, {Machado} P, {Malaguti}
  G, {Min} M, {N{\o}rgaard-Nielsen} HU, {Rataj} M, {Ray} T, {Ribas} I, {Swain}
  M, {Szabo} R, {Werner} S, {Barstow} J, {Burleigh} M, {Cho} J, {du Foresto}
  VC, {Coustenis} A, {Decin} L, {Encrenaz} T, {Galand } M, {Gillon} M, {Helled}
  R, {Morales} JC, {Mu{\~n}oz} AG, {Moneti} A, {Pagano} I, {Pascale} E,
  {Piccioni} G, {Pinfield} D, {Sarkar} S, {Selsis} F, {Tennyson} J, {Triaud} A,
  {Venot} O, {Waldmann} I, {Waltham} D, {Wright} G, {Amiaux} J, {Augu{\`e}res}
  JL, {Berth{\'e}} M, {Bezawada} N, {Bishop} G, {Bowles} N, {Coffey} D,
  {Colom{\'e}} J, {Crook} M, {Crouzet} PE, {Da Peppo} V, {Sanz} IE, {Focardi}
  M, {Frericks} M, {Hunt} T, {Kohley} R, {Middleton} K, {Morgante} G,
  {Ottensamer} R, {Pace} E, {Pearson} C, {Stamper} R, {Symonds} K, {Rengel} M,
  {Renotte} E, {Ade} P, {Affer} L, {Alard} C, {Allard} N, {Altieri} F,
  {Andr{\'e}} Y, {Arena} C, {Argyriou} I, {Aylward} A, {Baccani} C, {Bakos} G,
  {Banaszkiewicz} M, {Barlow} M, {Batista} V, {Bellucci} G, {Benatti} S,
  {Bernardi} P, {B{\'e}zard} B, {Blecka} M, {Bolmont} E, {Bonfond} B, {Bonito}
  R, {Bonomo} AS, {Brucato} JR, {Brun} AS, {Bryson} I, {Bujwan} W, {Casewell}
  S, {Charnay} B, {Pestellini} CC, {Chen} G, {Ciaravella} A, {Claudi} R,
  {Cl{\'e}dassou} R, {Damasso} M, {Damiano} M, {Danielski} C, {Deroo} P, {Di
  Giorgio} AM, {Dominik} C, {Doublier} V, {Doyle} S, {Doyon} R, {Drummond} B,
  {Duong} B, {Eales} S, {Edwards} B, {Farina} M, {Flaccomio} E, {Fletcher} L,
  {Forget} F, {Fossey} S, {Fr{\"a}nz} M, {Fujii} Y, {Garc{\'\i}a-Piquer} {\'A},
  {Gear} W, {Geoffray} H, {G{\'e}rard} JC, {Gesa} L, {Gomez} H, {Graczyk} R,
  {Griffith} C, {Grodent} D, {Guarcello} MG, {Gustin} J, {Hamano} K, {Hargrave}
  P, {Hello} Y, {Heng} K, {Herrero} E, {Hornstrup} A, {Hubert} B, {Ida} S,
  {Ikoma} M, {Iro} N, {Irwin} P, {Jarchow} C, {Jaubert} J, {Jones} H, {Julien}
  Q, {Kameda} S, {Kerschbaum} F, {Kervella} P, {Koskinen} T, {Krijger} M,
  {Krupp} N, {Lafarga} M, {Landini} F, {Lellouch} E, {Leto} G, {Luntzer} A,
  {Rank-L{\"u}ftinger} T, {Maggio} A, {Maldonado} J, {Maillard} JP, {Mall} U,
  {Marquette} JB, {Mathis} S, {Maxted} P, {Matsuo} T, {Medvedev} A, {Miguel} Y,
  {Minier} V, {Morello} G, {Mura} A, {Narita} N, {Nascimbeni} V, {Nguyen Tong}
  N, {Noce} V, {Oliva} F, {Palle} E, {Palmer} P, {Pancrazzi} M, {Papageorgiou}
  A, {Parmentier} V, {Perger} M, {Petralia} A, {Pezzuto} S, {Pierrehumbert} R,
  {Pillitteri} I, {Piotto} G, {Pisano} G, {Prisinzano} L, {Radioti} A,
  {R{\'e}ess} JM, {Rezac} L, {Rocchetto} M, {Rosich} A, {Sanna} N, {Santerne}
  A, {Savini} G, {Scandariato} G, {Sicardy} B, {Sierra} C, {Sindoni} G, {Skup}
  K, {Snellen} I, {Sobiecki} M, {Soret} L, {Sozzetti} A, {Stiepen} A,
  {Strugarek} A, {Taylor} J, {Taylor} W, {Terenzi} L, {Tessenyi} M, {Tsiaras}
  A, {Tucker} C, {Valencia} D, {Vasisht} G, {Vazan} A, {Vilardell} F,
  {Vinatier} S, {Viti} S, {Waters} R, {Wawer} P, {Wawrzaszek} A, {Whitworth} A,
  {Yung} YL, {Yurchenko} SN, {Osorio} MRZ, {Zellem} R, {Zingales} T, {Zwart} F
  (2018) {A chemical survey of exoplanets with ARIEL}. Experimental Astronomy
  46(1):135--209, \doi{10.1007/s10686-018-9598-x}

\bibitem[{Visser and Bergin(2012)}]{VisserANDBergin2012}
Visser R, Bergin EA (2012) Fundamental aspects of episodic accretion chemistry
  explored with single-point models. The Astrophysical Journal Letters
  754(1):18, \doi{10.1088/2041-8205/754/1/L18}

\bibitem[{Visser et~al.(2015)Visser, Bergin, and Jorgensen}]{Visser2015}
Visser R, Bergin EA, Jorgensen JK (2015) Chemical tracers of episodic accretion
  in low-mass protostars. Astronomy \& Astrophysics 577:12,
  \doi{10.1051/0004-6361/201425365}

\bibitem[{Vorobyov(2013)}]{Vorobyov2013}
Vorobyov EI (2013) Formation of giant planets and brown dwarfs on wide orbits.
  Astronomy \& Astrophysics 552:15, \doi{10.1051/0004-6361/201220601}

\bibitem[{{Williams} and {Cieza}(2011)}]{Williams2011}
{Williams} JP, {Cieza} LA (2011) {Protoplanetary Disks and Their Evolution}.
  \araa 49(1):67--117, \doi{10.1146/annurev-astro-081710-102548},
  \eprint{1103.0556}

\bibitem[{{Wyatt}(2008)}]{Wyatt2008}
{Wyatt} MC (2008) {Evolution of debris disks.} \araa 46:339--383,
  \doi{10.1146/annurev.astro.45.051806.110525}

\end{thebibliography}

%
%

\end{document}